\PassOptionsToPackage{table}{xcolor}
\documentclass[sigconf]{acmart}

\copyrightyear{2026}
\acmYear{2026}
\setcopyright{cc}
\setcctype{by}
\acmConference[SIGCSE TS 2026]{Proceedings of the 57th ACM Technical Symposium on Computer Science Education V.2}{February 18--21, 2026}{St. Louis, MO, USA}
\acmBooktitle{Proceedings of the 57th ACM Technical Symposium on Computer Science Education V.2 (SIGCSE TS 2026), February 18--21, 2026, St. Louis, MO, USA}
\acmPrice{}
\acmDOI{10.1145/3770761.3777323}
\acmISBN{979-8-4007-2255-4/2026/02}

\begin{document}

\title{Automated Program Repair of Uncompilable Student Code}

\author{Griffin Pitts}
\email{wgpitts@ncsu.edu}
\orcid{0009-0004-3111-6118}
\affiliation{
  \institution{North Carolina State University}
  \city{Raleigh}
  \state{North Carolina}
  \country{USA}
}
\author{Aum Pandya}
\authornote{All authors contributed equally to this research.}
\email{apandya4@ncsu.edu}
\orcid{0000-0002-2306-3464}
\affiliation{
  \institution{North Carolina State University}
  \city{Raleigh}
  \state{North Carolina}
  \country{USA}
}
\author{Darsh Rank}
\authornotemark[1]
\email{drank@ncsu.edu}
\orcid{0009-0008-2482-1731}
\affiliation{
  \institution{North Carolina State University}
  \city{Raleigh}
  \state{North Carolina}
  \country{USA}
}

\author{Tirth Bhatt}
\email{tjbhatt@ncsu.edu}
\orcid{0009-0001-5139-3081}
\affiliation{
  \institution{North Carolina State University}
  \city{Raleigh}
  \state{North Carolina}
  \country{USA}
}
\author{Muntasir Hoq}
\email{mhoq@ncsu.edu}
\orcid{0000-0003-2591-0476}
\affiliation{
  \institution{North Carolina State University}
  \city{Raleigh}
  \state{North Carolina}
  \country{USA}
}
\author{Bita Akram}
\email{bakram@ncsu.edu}
\orcid{0000-0001-5195-5841}
\affiliation{
  \institution{North Carolina State University}
  \city{Raleigh}
  \state{North Carolina}
  \country{USA}
}

\begin{abstract}
    A significant portion of student programming submissions in CS1 learning environments are uncompilable, limiting their use in student modeling and downstream knowledge tracing. Traditional modeling pipelines often exclude these cases, discarding observations of student learning. This study investigates automated program repair as a strategy to recover uncompilable code while preserving students’ structural intent for use in student modeling. Within this framework, we assess large language models (LLMs) as repair agents under high- and low-context prompting conditions. Repairs were evaluated for compilability, edit distance, and preservation of students' original structure and logic. While all models produced compilable repairs, they differed in how well they preserve students’ control flow and code structure, affecting their pedagogical utility. By recovering uncompilable submissions, this work enables richer and more comprehensive analyses of learners’ coding processes and development over time.
\end{abstract}

\maketitle


\section{Introduction}

Intelligent tutoring systems and related student modeling approaches often rely on large volumes of student submissions to track learning and provide personalized support. Yet, a significant share of novice programs are frequently uncompilable due to syntax errors. These uncompilable submissions are often excluded because they lack an evaluable score or performance measure, with existing methods dependent on code that can be parsed or executed, such as for generating abstract syntax trees in structural analyses \cite{hoq2025pattern}. However, uncompilable code can contain important information about students’ intermediate reasoning and knowledge states, informing more complete models of their learning. Excluding such data reduces coverage and discards potentially meaningful evidence of student understanding.

Automated Program Repair (APR) has been widely studied in software engineering and computing education as a means to fix erroneous code. In educational contexts, APR is used to provide students with repaired solutions or feedback on logical errors. Early systems such as DeepFix demonstrated the feasibility of large-scale repair using neural sequence models \cite{gupta2017deepfix}, while Koutcheme et al. \cite{koutcheme2023automated} explored the use of large language models for code infilling, applying generative approaches to repair student programs and provide feedback on functional or semantic errors. Later work such as CEMR \cite{wan2024automated} further advanced semantic repair by learning contextual edit operations from student submissions. While APR approaches are effective for producing correct code, they can also overwrite the student’s original approach. In the context of modeling students’ knowledge for feedback and analysis, semantic rewrites risk distorting the very evidence of learning we seek to model.

Syntax-only repair offers a more targeted opportunity. Novice submissions often fail due to small surface-level mistakes such as missing semicolons, unmatched braces, or undeclared variables that can be resolved with minimal edits. Lightweight fixes can retain the student’s original structure while making the code executable, thereby preserving the integrity of their learning trajectory. Prior work has explored such corrections through rule-based systems, for instance Takhar and Aggarwal \cite{takhar2019grading}, who inserted missing tokens to make student programs compilable, though their approach was limited to a narrow set of predefined error patterns. Extending this line of research, we explore how large language models perform syntax-only repair on uncompilable student code, examining how model type and prompting context influence repair outcomes.

\section{Methodology}

\subsection{Dataset}
We use a publicly available dataset sourced from the CodeWorkout platform~\cite{edwards2017codeworkout}, an online programming learning environment. CodeWorkout provides short, auto-graded Java exercises relating to topics such as logic, conditionals, loops, and arrays. The dataset contains $57{,}670$ anonymized Java submissions from $368$ students in a CS1 course offered at a U.S. university during the Spring $2019$ semester. Of these submissions, $18{,}787$ submissions were correct and $38{,}883$ were incorrect. Among the incorrect submissions, $9{,}906$ (approximately 25\%) failed to compile.

\subsection{Evaluation}

To capture a representative range of novice errors, two problems were randomly selected from a pool of 50 in the CodeWorkout dataset. The selected problems addressed logic, conditionals, loops, arrays, and string manipulation. From these, 100 uncompilable Java submissions were randomly sampled. Each submission was then processed under six experimental conditions, combining three language models (GPT-5, Claude 3.5 Haiku, and Gemini 2.5 Flash) with two prompting contexts (low and high), yielding a total of 600 repaired outputs for evaluation. In the low-context condition, models received only the uncompilable student code and instructions to perform syntax-only repair with minimal edits that preserved control flow, identifiers, and formatting. The high-context condition additionally provided the compiler message, problem statement, and few-shot examples of correct and incorrect repairs.

\subsubsection*{Evaluation Criteria}
Repair quality was assessed using compilation success, edit distance, and human evaluation. Compilation success indicated whether the repaired code executed without syntax errors. Edit distance, computed using normalized Levenshtein distance, measured how closely the repair aligned with the original code. 

For human evaluation, four experts independently annotated all repaired outputs for Structural Preservation (SP) and Logical Preservation (LP). SP was coded as 1 if the repaired code maintained the original control flow and 0 otherwise. LP was coded as 1 if the repair consisted only of syntactic edits, such as adding missing delimiters or correcting variable names, and 0 if the repair changed the logical or semantic structure of the code. 

The annotation process was iterative: experts coded an initial subset ($10$\% of the repaired programs) jointly to align on definitions, then proceeded independently, calculating Cohen’s Kappa ($\kappa$) after each round. When agreement fell below 0.80, in line with \cite{landis1977measurement}, disagreements were resolved and the codebook was refined. Inter-rater agreement exceeding $\kappa$ = 0.80 was achieved on the second round, after which the full set was coded and analysis conducted. Statistical analyses included chi-square tests of independence for categorical outcomes (compilation success, SP, and LP) and ANOVA for continuous measures (edit distance).

\section{Results}

Our aim was to examine how different large language models and prompting contexts influence the quality of syntax-only code repair. Compilation success rates were high and consistent across models, with GPT-5 achieving 98.5\% (591/600), Claude 3.5 96\% (576/600), and Gemini 2.5 95.5\% (573/600) compilable repairs. There were no statistically significant differences in compilation rates among the models ($\chi^2$(2, N = 600) = 3.21, p = .201). Prompting condition also did not significantly affect compilability ($\chi^2$(1, N = 600) = 0.21, p = 0.649), indicating that all three models effectively produced compilable repairs even under low-context prompting.

Edit distance significantly differed by model ($F$(2, 594) = 16.22, $p$ < 0.001), with GPT-5 producing the smallest average edits ($11.4$), followed by Gemini~2.5 ($13.8$) and Claude~3.5 ($24.4$). Prompting condition had no significant effect on edit distance ($F$(1, 594) = 0.004, $p$ = 0.95), contrary to our hypothesis that providing additional context might encourage models to over-correct or attempt to solve the problem rather than perform syntax-only repairs.

Based on human evaluation, Structural Preservation (SP) differed by model ($\chi^2$(2, N = 579) = 18.10, $p$ < 0.001), with GPT-5 maintaining control flow in 190 of 197 repairs (96.4\%), compared to Gemini~2.5 (186 of 190; 97.9\%) and Claude~3.5 (170 of 192; 88.5\%). Logic Preservation (LP) also differed significantly ($\chi^2$(2, N = 549) = 22.36, $p$ < 0.001), with GPT-5 showing the highest proportion of logic-preserving repairs (166 of 192; 86.5\%), followed by Gemini~2.5 (156 of 186; 83.9\%) and Claude~3.5 (116 of 171; 67.8\%). Prompting condition had no significant effect on either measure (SP: $\chi^2$(1, N = 579) = 0.02, $p$ = 0.88; LP: $\chi^2$(1, N = 549) = 0.00, $p$ = 0.97).

\section{Discussion \& Future Work}
Our findings indicate that large language models can reliably perform repairs on short student code snippets, often conforming well to the explicit prompt instructions. However, similar to the tendencies observed by Řechtáčková et al. \cite{vrechtavckova2025finding}, the models occasionally strayed from the intended pedagogical scope by making stylistic or structural edits that went beyond minimal correction. This pattern reflects an ongoing challenge in achieving pedagogical alignment with LLMs used in educational settings \cite{pitts2025surveyllmbasedapplicationsprogramming}.

While the present evaluation focused on compilation outcomes, edit distance, and expert annotation, more rigorous and fine-grained evaluation frameworks are needed to capture the pedagogical quality and instructional appropriateness of model-generated repairs. Future studies should incorporate larger and more diverse code samples and investigate how repaired submissions can be integrated into student modeling and intelligent tutoring system pipelines to better represent students’ intermediate reasoning, misconceptions, and problem-solving progress. Further work should also consider pedagogical alignment more broadly, exploring how fine-tuning, context conditioning, or data augmentation can better align LLM behavior with the instructional tasks it is expected to perform.

\begin{acks} 
This research was supported by NSF under Grant \#2418658. Any opinions, findings, and conclusions expressed in this material are those of the authors and do not necessarily reflect the views of NSF.
\end{acks}

\bibliographystyle{ACM-Reference-Format}
\balance
\bibliography{ref}

@article{hoq2025pattern,
  title={Pattern-based Knowledge Component Extraction from Student Code Using Representation Learning},
  author={Hoq, Muntasir and Pitts, Griffin and Lan, Andrew and Brusilovsky, Peter and Akram, Bita},
  journal={arXiv:2508.09281},
  year={2025}
}

@article{pitts2025surveyllmbasedapplicationsprogramming,
      title={A Survey of LLM-Based Applications in Programming Education: Balancing Automation and Human Oversight}, 
      author={Pitts, Griffin and Hridi, Anurata Prabha and Lekshmi-Narayanan, Arun-Balajiee},
      journal={arXiv preprint arXiv:2510.03719},
      year={2025}
}

@inproceedings{gupta2017deepfix,
  title={Deepfix: Fixing common c language errors by deep learning},
  author={Gupta, Rahul and Pal, Soham and Kanade, Aditya and Shevade, Shirish},
  booktitle={AAAI},
  volume={31},
  number={1},
  year={2017}
}

@inproceedings{takhar2019grading,
  title={Grading uncompilable programs},
  author={Takhar, Rohit and Aggarwal, Varun},
  booktitle={AAAI},
  volume={33},
  number={01},
  pages={9389--9396},
  year={2019}
}

@inproceedings{vrechtavckova2025finding,
  title={Finding misleading identifiers in novice code using LLMs},
  author={{\v{R}}echt{\'a}{\v{c}}kov{\'a}, Anna and Maximova, Alexandra and Pitts, Griffin},
  booktitle={Proceedings of the 56th ACM Technical Symposium on Computer Science Education V. 2},
  pages={1595--1596},
  year={2025}
}

@article{landis1977measurement,
  title={The measurement of observer agreement for categorical data},
  author={Landis, J Richard and Koch, Gary G},
  journal={biometrics},
  pages={159--174},
  year={1977},
  publisher={JSTOR}
}

@inproceedings{edwards2017codeworkout,
  title={CodeWorkout: short programming exercises with built-in data collection},
  author={Edwards, Stephen H and Murali, Krishnan Panamalai},
  booktitle={Proceedings of the 2017 ACM conference on innovation and technology in computer science education},
  year={2017}
}

@inproceedings{koutcheme2023automated,
  title={Automated program repair using generative models for code infilling},
  author={Koutcheme, Charles and Sarsa, Sami and Leinonen, Juho and Hellas, Arto and Denny, Paul},
  booktitle={International Conference on Artificial Intelligence in Education},
  pages={798--803},
  year={2023},
  organization={Springer}
}

@article{wan2024automated,
  title={Automated program repair for introductory programming assignments},
  author={Wan, Han and Luo, Hongzhen and Li, Mengying and Luo, Xiaoyan},
  journal={IEEE ToLT},
  year={2024},
  publisher={IEEE}
}

\end{document}